# Current Leads for Superconducting Magnets of ADS Injector I

WANG Bing(王冰), PENG Quan-ling(彭全岭), YANG Xiang-chen (杨向臣),
Li Shao-peng(李少鹏)
Institute of High Energy Physics, Chinese Academy of Sciences, Beijing 100049, China

**Abstract**

In ADS Injector I, there are six superconducting magnets in each cryomodule. Each superconducting magnet contains a solenoid magnet, a horizontal dipole corrector (HDC) and a vertical dipole corrector (VDC). Six current leads will be required for powering the electrical circuits, from room temperature to the 2.1K liquid helium bath. Two leads carry 100A current for the solenoid magnet while the other four carry 12A for the HDC and the VDC. This paper presents the principle for current lead optimization, which includes the cooling methods, the choice of material and structure, and the issues for current lead integration as well.

## 1. Introduction

ADS is the abbreviation for Accelerator Driven Sub-critical System. The main purpose of developing ADS lies in separating and transmuting irradiated nuclear fuel [1]. Three major units constitute the separation and transmutation system, i.e., the accelerator unit, target unit, and the separation unit. The schematic layout for ADS injection I is shown in Figure 1. The detailed cryomodule drawing as shown in Figure 1 demonstrates 6 superconducting spoke cavities, 5 superconducting magnets and 5 beam position monitors connected in series along the beam line. The proton beam energy will be upgraded from inlet 3Mev to outlet 5Mev by these 6 cavities. The 5 superconducting magnets, on the other hand, will provide the beam focusing and orbit correction. As an important component, the superconducting magnet prototype have been perfectly designed and passed the vertical test successfully in Nov. 2012 [2].

Current leads are applied to deliver room-temperature electrical power to cryogenic superconducting magnets that are immersed in liquid helium. The temperature difference between the two ends and joule heat produced by the excited current make the current leads the dominant heat source into the cryogenic system. To simplify the cryogenic system, the superconducting magnet will use the same state liquid helium for the spoke cavity, liquid helium will sink into the magnet by its own gravity. Conduction cooled current leads are chosen because the negative pressure of 30mbar, 2.1K liquid helium can not provide enough pressure as the traditional gas cooled current leads. Unlike the traditional conduction-cooled current leads, the current lead that will be used consists of a pure copper coated brass rod, where the brass core has a constant diameter while the thickness of the coated pure copper varies along the length. Two stage thermal anchors, one connected to 5K helium gas shield, another connected to the 80K liquid nitrogen shield, are installed on the leads to intercept the heat leak into the cryogenic system in step.

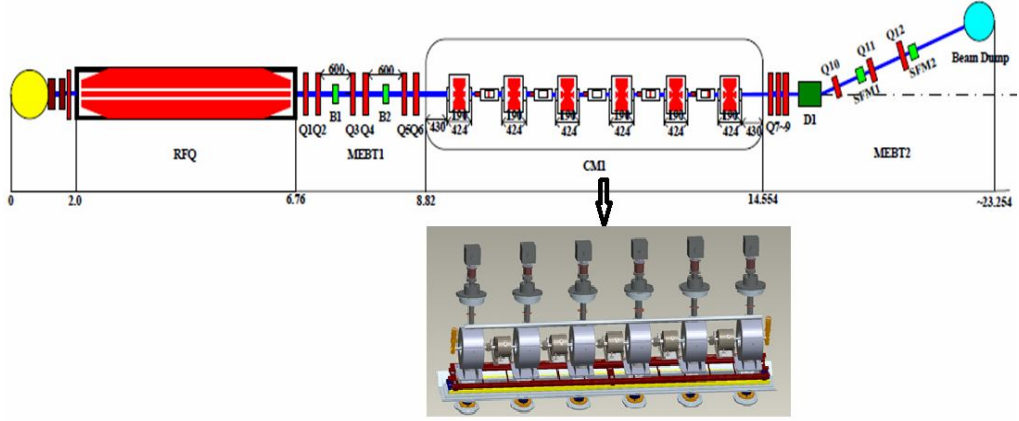

Fig. 1. Schematic layout for ADS injection I and the detailed cryomodule

2. Conduction-cooled current leads

The superconducting magnets and the spoke cavities will be immersed in 30mbar, 2.1K liquid helium bath, conduction cooled leads will be chosen for the magnets. The calculate cases here used for the lead optimization will be separated into three temperature segments, from 300K to 80K, from 80K to 5K and from 5K to 2K.

Fig 2 shows a current lead with the length of L, cross section area of A, carrying current of I. T0 and T1 represent the temperature of the cold and warm end. Suppose the lead has uniform temperature distribution on cross section, $\lambda(T)$ and $\rho(T)$ are the thermal conductivity and the electric resistivity of the material, Q0 is the heat flux at point 0.

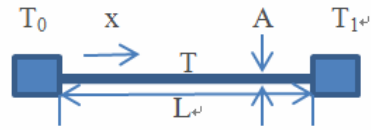

Fig 2. Schematic layout of a conduction current lead

Consider a short length current lead from the cold point T0 to point x, the heat flux Q0 includes electric dissipation term and conduction heat term, which can be written as [3]:

$$Q_0 = A\lambda(T)\frac{dT}{dx} + \frac{I^2}{A}\int_0^x \rho(x)dx \quad (1)$$

The optimized length x is in that situation when the heat flux Q0 gets minimum value, which should meet:

$$\rho\frac{I^2}{A} + A\frac{d}{dx}(\lambda\frac{dT}{dx}) = 0 \quad (2)$$

Set u=dT/dx, equation (2) can be integrated as:

$$u = \frac{dT}{dx} = \frac{B}{\lambda} - \frac{I^2}{\lambda A^2}x \quad (3)$$

where B is the integral constant. Then we have

$$u^2 = \left(\frac{dT}{dx}\right)^2 = \frac{I^2}{A^2\lambda^2}\left[C - 2\int_{T_0}^{T} \rho\,\lambda\,dT\right] \tag{4}$$

Noted that, the second order small amount is omitted here, the driven constant $C=B2A2/I2$, also $x=dx=\lambda/BdT$ have used here when x near equal to 0. Constant C can be got from (4) and (1) for x=0 and T=T0, which is

$$Q_0 = A\lambda(T_0)\left(\frac{dT}{dx}\right)_0 = I\sqrt{C} \tag{5}$$

The optimized factor for the current lead can be written as

$$\frac{L}{A} = \frac{1}{I}\int_{T_0}^{T_1} \frac{\lambda}{\sqrt{C - 2\int_{T_0}^{T}\rho\,\lambda\,dT}}dT \tag{6}$$

It can be used to optimize length of the current lead for a given current and cross section area of A. In equation (6), constant C must meet

$$C \geq 2\int_{T_0}^{T}\rho\,\lambda\,dT \tag{7}$$

From (5), the minimum heat low temperature side is

$$(Q_0)_{min} = I\sqrt{2\int_{T_0}^{T_1}\rho\,\lambda\,dT} \tag{8}$$

Then the optimized factor for the current lead can be rewritten as

$$\frac{L}{A} = \frac{1}{I}\int_{T_0}^{T_1}\frac{\lambda}{\sqrt{2\int_{T}^{T_1}\rho\,\lambda\,dT}}dT \tag{9}$$

For pure metals, the thermal conductivity $\lambda(T)$ and the electric resistivity $\rho(T)$ are inverse related, that meet Wiedemann-Franz law

$$k(T)\rho(T) = L_0 T \tag{10}$$

Here, $L_0 = 2.45\times 10^{-8}\,W\,\Omega\,K^{-2}$. Then the above equation (8) can be rewritten as

$$(Q_0)_{min} = I\sqrt{T_1^2 - T_0^2} \tag{11}$$

When optimize the current length, equation (11) can be rewritten as

$$\frac{L}{A} = \frac{1}{I}\int_{T_0}^{T_1}\frac{\lambda}{\sqrt{T_1^2 - T^2}}dT \tag{12}$$

### 3. ADS current leads concept design

Pure copper are widely accepted as a good conductor of electricity and its $\rho$ small electric resistivity and high conductivity in comparison with many other metals, but the use of high purity materials tends to make the leads unstable and liable to burn out at currents only slightly above the optimum [2]. To achieve low heat load to the cryogenic system and guarantee the stability of the lead, hybrid conductor current leads will be chosen, as that of CERN LHC dipole correctors and DESY XFEL superconducting magnets [4]. The hybrid conductor is a copper plated brass rod: the brass has a fixed diameter of 4mm while difference thickness of coated pure copper can be selected in the required section in order to adjust the total length of the current leads as illustrated in (9) and (12). Fig 3 is a sketch of a current lead with pure copper coated on a brass core. The pure copper has RRR of 120 while brass has RRR only of 5. The electric resistivity and thermal conductivity properties at different temperature for pure copper and brass core are revealed in Fig 4. It can be seen from the figure, at any case, whether at high or at low temperature, resistivity of the pure copper is about two orders lower than that of the brass copper, and carries almost all of the loaded current. At the same time, the coated pure copper has a high heat conductivity whether at high and low temperature compare with that of the brass copper, so the pure is the main conduction heat source to the low temperature side. While the brass copper core, which is a poorer thermal conductor with high resistivity, will serve both as a heat sink and a support for the pure copper to prevent of the current lead overheat in case of carrying excess current.

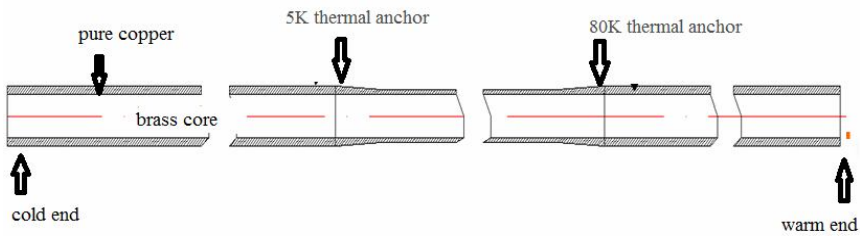

Fig. 3. Copper plated brass rod current lead

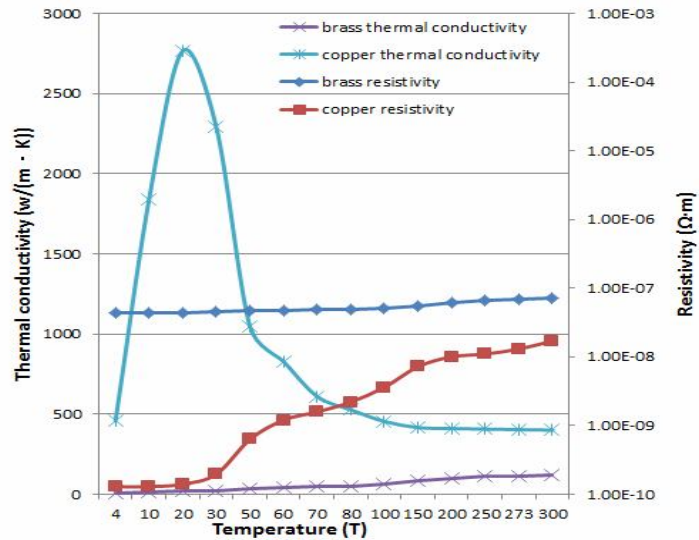

Fig. 4. Electric resistivity and thermal conductivity of pure copper and brass

AS shown in Fig. 3, for ADS superconducting solenoid magnet, two steps of thermal anchors, 80K and 5K, are designed to further intercept the heat leak towards the cryogenic system. The lead is optimized in each of the three segments, the temperature range changes from 300K to 80K, from 80K to 5K and from 5K to 2K. In order to reduce the heat to the 2K cryogenic system, the current lead was optimized at 100A from (8). The required length at each thermal section was optimized according to (12). The optimized results are listed in Table 1. Noted here that the temperature at each thermal anchor is higher than cold shield because there is temperature increase through a copper braiding that connect the cold shield to the thermal anchor. In 8~85K segment a smaller cross section coated pure copper is chosen in order to adjust the total length of current lead.

Table 1. Optimized results of 100A current leads (D1 and D2 represent the cross-sectional area of the brass core, before and after copper plated, respectively. Q100A is the heat load for the current lead at 100A. L is the length of current lead. )

|  | D1(mm) | D2(mm) | Q100A(W) | L(mm) |
|---|---|---|---|---|
| 2-8K | 4 | 5.65 | 0.2 | 600 |
| 8-85K | 4 | 4.89 | 1.68 | 600 |
| 85-300K | 4 | 5.65 | 1.83 | 500 |

**4. Engineering design**

The insulation of current leads is crucial and necessary. The lead is designed as an electrically insulated hybrid conductor contained in a thin stainless steel tube welded to the supporting flanges at the two extremities in 300K and 2K. The electrical insulation is made with a Kapton tube between the stainless steel tube and the lead. The cold end of the stainless steel tube is open to the liquid helium bath and static helium gas stratifies inside the tube. There is a small gas tank at the room temperature to guarantee the leak tightness between the helium gas and the outside environment.

To further intercept the heat leak towards the cryogenic system, 80K and 5K thermal anchors are made by connecting together the lead and the cryogenic lines that transporting liquid nitrogen or helium. First, the leads are clamped between two copper plates with grooves for housing the lead, and then, the copper plates are joined with the cryogenic transporting lines by copper woven straps. Fig 5 is 3D engineering drawing of current leads. With the participation of copper who has perfect electrical conductivity, this design assures good heat transfer from the tubes carrying cryogenic liquid to the leads, as a consequence, the temperatures at the place of thermal anchors are almost the same as or a little bit higher than that of cryogenic transporting lines. In view of the rigid space constraints imposed by the cryostat's configuration, the current leads are pre-shaped as required for the integration in the cryostat.

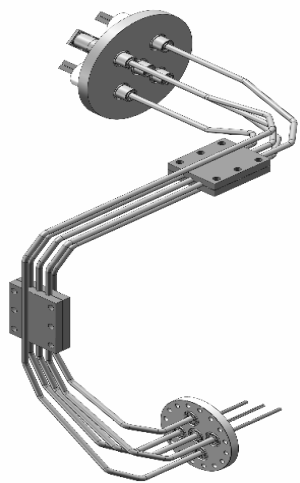

Fig 5. 3D engineering drawing of current lead

## 5. Conclusion

As an important component of superconducting magnets for ADS, current leads have been perfectly designed. In order to reduce the total heat load to the cryogenic system, the operation current for the solenoid is 100A, maximum current for the corrector coils are 20A. Lengths of the current leads are optimized according to the cross section area for the three segments between the thermal anchors.